\documentstyle[aps,prb,twocolumn,epsf,floats]{revtex}
\draft
\newcommand{\ro}{\bbox{\rho}}
\newcommand{\bpi}{\bbox{\pi}}
\newcommand{\bPi}{\bbox{\Pi}}
\newcommand{\bn}{\bbox{\nabla}}
\begin{document}
\wideabs{
\title{Charged hydrogenic problem in a magnetic field:
    Non-commutative translations, unitary transformations,
    and coherent states}
\author{A. B. Dzyubenko\cite{ABD}}
\address{Department of Physics, University at Buffalo, SUNY,
          Buffalo, NY 14260, USA}
\date{\today}
\maketitle
\begin{abstract}
An operator formalism is developed for a description of charged
electron-hole complexes in magnetic fields.
A novel unitary transformation of the Hamiltonian that allows one
to partially separate the center-of-mass and internal motions
is proposed.
We study the operator algebra that leads to the appearance
of new effective particles, electrons and holes with modified
interparticle interactions, and their coherent states in magnetic fields.
The developed formalism is used for studying
a two-dimensional negatively charged magnetoexciton $X^-$.
It is shown that Fano-resonances are
present in the spectra of internal $X^-$ transitions,
indicating the existence of three-particle quasi-bound states
embedded in the continuum of higher Landau levels.
\end{abstract}
\pacs{73.20.Mf,71.35.Ji,73.43.Lp}
}

\section{Introduction}

A quantum mechanical description of a system of charged
interacting particles in a magnetic field has long played
a central role in many solid state \cite{Kohn,Knox,Brown,G&D} and
atomic \cite{Lamb,Simon,Hirsch} physics problems.
Recently, there has been considerable interest in such
problems in the context of charged collective
excitations in a two-dimensional electron gas
in strong magnetic fields, \cite{coll}
excitations in the fractional quantum Hall effect, \cite{Girvin,Ezawa}
charged skyrmions, \cite{Pasquier}
and charged magnetoexcitons in quantum wells. \cite{X-th}
The internal and the center-of-mass (CM) motions
are generally coupled in a magnetic field $B$.
Systems with constant charge-to-mass ratio,
such as one-component electron systems, \cite{Kohn,Simon,Hirsch}
are the exception.
For a neutral problem, such as the two-body hydrogen atom \cite{Lamb}
or the exciton, \cite{Knox,G&D} there exists a possibility to separate
the CM and internal variables in the Schr\"odinger equation.
Generally, however, only a partial separation is possible. \cite{Simon}
Several formalisms have been developed in order to perform
such a separation in a magnetic field. \cite{Hirsch}

In this work, we propose a new operator approach for
charged electron-hole systems in a magnetic field.
This approach is a development of Ref.~\onlinecite{SSC},
which has exploited an exact dynamical symmetry, the non-commutative
magnetic translations (MT). \cite{Brown,Simon,Hirsch,Ezawa}
Here we show that in order to maintain both the MT and
axial symmetry about the ${\bf B}$-axis, one can use a description
in terms of {\em coherent states of new effective particles},
electrons ($e$) and holes ($h$) in $B$
or, alternatively, perform
{\em a unitary transformation of the Hamiltonian}.
The interparticle $e$--$h$ interaction is modified by the transformation.
We show that closed analytic expressions can be found for
matrix elements of the new interaction after summing
contributions from an infinite number of higher Landau levels (LL's).

  The developed formalism is applied in this work to a description
of a two-dimensional (2D) charged magnetoexciton $X^{-}$,
a bound state of two electrons and one hole, in higher LL's.
Charged magnetoexcitons have recently been extensively studied
experimentally \cite{X-exp} and theoretically. \cite{X-th,Dz&S_PRL}
Spectral properties of a three-body problem in a magnetic field
present a considerable general theoretical interest. \cite{Simon,Hirsch}
Our approach is capable of describing the interaction between
discrete $X^-$ states and the three-particle $2e$--$h$ continuum.
We demonstrate that Fano-resonances \cite{Fano} are
present in the spectra of $X^-$ optical transitions
in strong fields. This is an indication that three-particle
resonances --- quasi-bound $X^-$ states embedded in a continuum ---
exist in 2D systems in higher LL's.

\section{Charged \lowercase{$e$}--\lowercase{$h$}
             systems in magnetic fields}
                 \label{sec:form}

We start with a short overview of the dynamical symmetries
of the Hamiltonian and of the operator formalism
that is most suitable for describing these symmetries
for single-particle \cite{Simon,Hirsch,Ezawa}
and few-particle \cite{SSC} $e$--$h$ states in $B$.

\subsection{Hamiltonian and dynamical symmetries}
                 \label{subsec:symmetry}

The Hamiltonian describing charged interacting 2D particles
in a perpendicular magnetic field ${\bf B}=(0,0,B)$ is
\begin{equation}
                \label{calH} 
  {\cal H} = \sum_{j} \frac{\hat{\bPi}_{j}^2}{2m_j} +
      \frac{1}{2} \sum_{i \neq j } U_{ij}( {\bf r}_i - {\bf r}_j )  \, ,
\end{equation}
where $\hat{\bPi}_j = -i\hbar \bn_j -  \frac{e_j}{c} {\bf A}({\bf r}_j)$
are kinematic momentum operators and
$U_{ij}({\bf r})$ are interaction potentials that can be arbitrary.
In the symmetric gauge
${\bf A} = \frac12 {\bf B} \times {\bf r}$,
the Hamiltonian ${\cal H}$ is characterized by {\em both\/}
axial symmetry, $[{\cal H},\hat{L}_z]=0$,
and by translational symmetry, $[{\cal H},\hat{\bf K}]=0$.
Here $\hat{L}_z=\sum_j ({\bf r}_j \times -i\hbar\bn_j)_z$
is the operator of the total angular momentum projection and
$\hat{\bf K} = \sum_{j} \hat{\bf K}_j$ is the MT
operator. \cite{Brown,Simon,Hirsch,Dz&S_PRL}
The generators of MT for individual particles are given by
$\hat{\bf K}_j = \hat{\bPi}_j - \frac{e_j}{c} {\bf r}_j \times {\bf B}$;
in the symmetric gauge,
$\hat{\bf K}_j({\bf B}) = \hat{\bbox \Pi}_j(-{\bf B})$.
Independent of the gauge, $\hat{\bf K}_j$ and $\hat{\bbox \Pi}_j$ commute:
$[\hat{K}_{jp},\hat{\Pi}_{jq}]=0$, $p,q=x,y$.

Note that $\hat{L}_z$ and $\hat{\bf K}^2$ commute with each other,
$[\hat{L}_z,\hat{\bf K}^2]=0$, and both commute with the Hamiltonian ${\cal H}$.
Therefore, exact eigenstates of ${\cal H}$ can be simultaneously labeled by
the total angular momentum projection $M_z$,
an eigenvalue of $\hat{L}_z$, and by an eigenvalue of $\hat{\bf K}^2$.
The important feature of  $\hat{\bf K}$ is
the {\em non-commutativity\/} of its components in $B$:
$[\hat{K}_x, \hat{K}_y] = - i \frac{\hbar B}{c} Q$,
where $Q = \sum_j e_j$ is the total charge.
Introducing the dimensionless operator
$\hat{{\bf k}} = \sqrt{c/\hbar B |Q|} \, \hat{\bf K}$, which
has canonically conjugate components, one obtains the lowering and raising
Bose ladder operators for the whole system \cite{Simon,Hirsch,Dz&S_PRL}
\begin{equation}
  \label{kmin}
  \hat{k}_{\pm} = \pm \frac{i}{\sqrt{2}} (\hat{k}_x  \pm i \hat{k}_y)
            \quad , \quad
    [\hat{k}_{+}, \hat{k}_{-}]=-\frac{Q}{|Q|} \, .
\end{equation}
From (\ref{kmin}) it follows that
$\hat{{\bf k}}^2 = \hat{k}_{+} \hat{k}_{-} + \hat{k}_{-} \hat{k}_{+}$
has the discrete oscillator eigenvalues $2k+1$, $k=0, 1, \ldots$.
There is a macroscopic (Landau) degeneracy in the oscillator
quantum number $k$, which qualitatively describes the center-of-rotation
of the charged complex in $B$.
Therefore, the exact eigenstates of ${\cal H}$ can be labeled by the
discrete quantum numbers $M_z$ and $k$;
for $e$--$h$ systems, because of the permutational symmetry,
there are additional exact quantum numbers,
the total spin of electrons, $S_e$, and holes, $S_h$,
and their projections, $S_{\rm ez}$ and $S_{\rm hz}$.
Degeneracy in $k$ leads to the existence of families of macroscopically
degenerate states. Because of the commutation relation
$[\hat{L}_z, \hat{k}_{\pm}]= \pm \hat{k}_{\pm}$,
the quantum numbers $M_z$ and $k$ are connected uniquely in each family;
this has been discussed in more detail elsewhere. \cite{Dz&S_PRL}

Another operator of interest is $\hat{\bPi} = \sum_{j} \hat{\bPi}_j$;
its components commute as
$[\hat{\Pi}_x, \hat{\Pi}_y] =  i \frac{\hbar B}{c} Q$.
In analogy with (\ref{kmin}), one therefore can introduce the second set
of raising and lowering Bose ladder operators
\begin{equation}
  \label{pimin}
 \hat{\pi}_{\pm} = \mp \frac{i}{\sqrt{2}} (\hat{\pi}_x  \pm i \hat{\pi}_y)
           \quad , \quad
    [\hat{\pi}_{+}, \hat{\pi}_{-}]= \frac{Q}{|Q|} \, ,
\end{equation}
where $\hat{\bpi} = \sqrt{c/\hbar B |Q|} \, \hat{\bPi}$.
Note, however, that the operators $\hat{\pi}_{\pm}$ do not commute
and, in general, do not form a simple algebra with the Hamiltonian.
A special case is when all particles have the same cyclotron
frequency $\omega_{\rm cj} = e_j B/m_j c$: the operator
algebra is closed
\begin{equation}
  \label{Pplus}
   [{\cal H}, \hat{\pi}_{\pm}] =
      \mp \hbar \omega_{\rm cj} \, \hat{\pi}_{\pm} \quad ,
        \quad  \frac{e_j}{m_j} = \mathrm{const}  \, ,
\end{equation}
and the CM and internal degrees of freedom separate in this case. \cite{Kohn}

\subsection{Single-particle $e$--$h$ states}
                  \label{subsec:single}

The formalism of Sec.~\ref{subsec:symmetry}
can be applied to non-interacting particles.
This leads to the description in terms of
so-called factored \cite{Simon,Hirsch,Ezawa,AHM_86}
single particle $e$- and $h$- states in a magnetic field
\begin{equation}
          \label{WF}
  \phi^{(e)}_{n m}({\bf r})=\phi^{(h)}_{n m}({\bf r})^{\ast} \, ,
\end{equation}
where $n$ is the LL number, which determines the energy
$\hbar\omega_{\rm ce(h)}(n+\case{1}{2})$, and
$\omega_{\rm ce(h)}=eB/m_{\rm e(h)}$ are the cyclotron frequencies.
The intra-LL oscillator quantum number is denoted here as $m$.
It is a single-particle version of $k$; analogously to $k$,
the energy is degenerate in $m$.
The wave functions (\ref{WF}) are constructed with the help
of the oscillator Bose ladder operators. \cite{Ezawa,AHM_86}
For electrons (charge $-e<0$)
\begin{equation}
              \label{e_LL}
\phi^{(e)}_{n m}({\bf r})=\frac{1}{\sqrt{n!m!}}
\langle {\bf r} |(A^{\dag}_{e})^n (B^{\dag}_{e})^m |0\rangle \, ,
\end{equation}
where the intra-LL operators
$B^{\dag}_{e}({\bf r}_j) =
 -i \sqrt{c/2 \hbar B e} \, (\hat{ K }_{jx} - i \hat{ K }_{jy})$
and the inter-LL operators
$A^{\dag}_{e}({\bf r}_j) =
 - i \sqrt{c/2 \hbar B e} \, (\hat{ \Pi }_{jx} + i \hat{ \Pi }_{jy})$
[cf.\ Eqs.~(\ref{kmin}) and (\ref{pimin})].
The operators commute as $[A_{e},A^{\dag}_{e}]=1$,
$[B_{e},B^{\dag}_{e}]=1$,
and $[A_{e},B^{\dag}_{e}]=[A_{e},B_{e}]=0$.
The analogous operators for the hole
(charge $e>0$) are
$B^{\dag}_{h}({\bf r}_h) =
 - i \sqrt{c/2 \hbar B e} \, (\hat{ K }_{hx} + i \hat{ K }_{hy})$
and
$A^{\dag}_{h}({\bf r}_h) =
  - i \sqrt{c/2 \hbar B e} \, (\hat{ \Pi }_{hx} - i \hat{ \Pi }_{hy})$;
we used the freedom of choosing an arbitrary phase of operators here.
These operators can be considered to be linear functions of
spatial coordinates and derivatives and have the form
\begin{eqnarray}
        \label{lad_en}
    A^{\dag}_{e}({\bf r}) = B^{\dag}_{h}({\bf r}) &=& \frac{1}{\sqrt{2}}
  \left( \frac{z}{2l_B} -
       2l_B \frac{\partial}{\partial z^{\ast}} \right) \, , \\
      \label{lad_em}
    B^{\dag}_{e}({\bf r}) = A^{\dag}_{h}({\bf r}) &=& \frac{1}{\sqrt{2}}
  \left( \frac{z^{\ast} }{2l_B} -
      2l_B \frac{\partial}{\partial z} \right) \, ,
\end{eqnarray}
$ z = x + iy$ is the 2D complex coordinate and
$l_B=(\hbar c/eB)^{1/2}$ is the magnetic length.
Single-particle angular momentum projection operators are
$\hat{L}_{\rm ze}=A^{\dag}_{e}A_{e} -  B^{\dag}_{e}B_{e}$
and $\hat{L}_{\rm zh}=B^{\dag}_{h}B_{h} - A^{\dag}_{h}A_{h}$,
so that $m_{\rm ze}=-m_{\rm zh}=n-m$.
For zero LL's, for example, the explicit form is
\begin{eqnarray}
              \label{zero_LL}
   & &\phi^{(e)}_{0m}({\bf r})^{\ast} = \phi^{(h)}_{0m}({\bf r}) \\
         \nonumber
   & & = \frac{1}{(2\pi m!l_B^2)^{1/2}}
   \left( \frac{z}{\sqrt{2}l_B} \right)^m
   \exp\left(-\frac{{\bf r}^2}{4l_B^2} \right)  \, .
\end{eqnarray}

\subsection{Three-particle $2e$--$h$ states: symmetries preserved}
                 \label{subsec:2e-h}

In what follows, we will consider the 2D three-particle
$2e$--$h$ states (the charged exciton $X^-$)
in a magnetic field ${\bf B}$.
The corresponding Hamiltonian is
${\cal H} = H_0 + H_{\rm int}$, where the free-particle part is given by
\begin{eqnarray}
                \label{H_0} 
  H_0 &=& \sum_{i=1,2} \frac{\hat{\bPi}_{ei}^2}{2m_e} +
                      \frac{\hat{\bPi}_{h}^2 }{2m_h}  \\
     \nonumber
   & \equiv &  \sum_{i=1,2} H_{0e}({\bf r}_i) + H_{0h}({\bf r}_h)  \, .
\end{eqnarray}
The interaction Hamiltonian is
\begin{eqnarray}
            \label{H_int} 
  H_{\rm int} &=& H_{\rm ee} + H_{\rm eh}  \, , \\
  H_{\rm ee} = U_{\rm ee}(|{\bf r}_1-{\bf r}_2|)
             \,  &,&  \,
  H_{\rm eh} = \sum_{i=1,2} U_{\rm eh}(|{\bf r}_i-{\bf r}_h|) \, .
\end{eqnarray}
In calculations (Sec.~\ref{sec:res}) we will consider
the Coulomb interaction $U_{\rm ee}=-U_{\rm eh}= e^2/\epsilon r$.
The total charge of the system $Q=-e<0$, and the
raising Bose operator is $\hat{k}_{-}$.
In terms of the {\em single-particle\/} Bose ladder operators
it takes the form
\begin{equation}
        \label{k-0}
     \hat{k}_{-} =
      B_e^{\dag}({\bf r}_1) + B_e^{\dag}({\bf r}_2) - B_h({\bf r}_h) \,
\end{equation}
and is a combination of creation and destruction operators.
The operator $\hat{k}_{-}$ is associated with the exact MT symmetry
and its diagonalization is a necessary step that allows one to keep
this symmetry intact.

It is convenient first to perform
an orthogonal transformation of the electron coordinates
$\{ {\bf r}_1,{\bf r}_2,{\bf r}_{h}\} \rightarrow
 \{ {\bf r},{\bf R}, {\bf r}_h \}$,
where ${\bf r} = ({\bf r}_1 - {\bf r}_2)/\sqrt{2}$, and
${\bf R} = ({\bf r}_1 + {\bf r}_2)/\sqrt{2}$
are the electron relative and CM coordinates.
The free Hamiltonian $H_0$ is a bilinear form in the coordinates
and spatial derivatives. Because of the orthogonality of the transformation,
$H_0$ conserves its form in the new  variables:
$H_0= H_{0e}({\bf r}) + H_{0e}({\bf R}) + H_{0h}({\bf r}_h)$.
The creation operator (\ref{kmin}) takes the form [cf.~(\ref{k-0})]
\begin{equation}
        \label{k-}
     \hat{k}_{-} = \sqrt{2}\,B_e^{\dag}({\bf R}) - B_h({\bf r}_h) \, .
\end{equation}
It can be diagonalized by introducing the transformed Bose
ladder operators \cite{SSC}
\begin{equation}
        \label{Bog1}
 \tilde{B}_e^{\dag}({\bf R}) \equiv
   u B^{\dag}_{e}({\bf R}) - v B_{h}({\bf r}_h)
 = \tilde{S} B_e^{\dag}({\bf R}) \tilde{S}^{\dag} \, ,
\end{equation}
This is the Bogoliubov canonical transformation generated by the
unitary operator (see, e.g., Refs.\ \onlinecite{Ezawa,Kirzhnits,Klauder})
\begin{eqnarray}
        \label{BogS}
      \tilde{S} & = &  \exp ( \Theta \tilde{\cal L} ) \, , \\
        \label{genL}
  \tilde{\cal L} & = &   B^{\dag}_{h}({\bf r}_h) B^{\dag}_{e}({\bf R})
           - B_e({\bf R})B_h({\bf r}_h) \, ,
\end{eqnarray}
where $\Theta$ is the transformation parameter and
$u= \cosh \Theta=\sqrt{2}$, $v=\sinh\Theta =1$.
Now we have $\hat{k}_- = \tilde{B}_e^{\dag}$ and
$\hat{{\bf k}}^2 = 2\tilde{B}_e^{\dag} \tilde{B}_e  + 1$.
The second linearly independent creation operator is
\begin{equation}
        \label{Bog2}
 \tilde{B}_h^{\dag}({\bf r}_h) =
            \tilde{S} B^{\dag}_{h}({\bf r}_h) \tilde{S}^{\dag} =
                      u B^{\dag}_{h}({\bf r}_h) - v B_{e}({\bf R})  \, ,
\end{equation}
Charged $e$--$h$ systems with an arbitrary number of particles
are considered in Appendix~\ref{Ap_A}.

We deal in fact with a sort of field-theoretical problem because
the number of relevant states is {\em infinite}. As an example,
the diagonalization of  $\hat{k}_{-}$ introduces a new vacuum state
\begin{equation}
          \label{S0}
 |0\rangle \stackrel {\hat{k}_{-}}{\longrightarrow}
            |\tilde{0} \rangle = \tilde{S} |0 \rangle  \, .
\end{equation}
A complete orthonormal basis compatible with
{\em both\/} axial and translational symmetries
can be constructed \cite{SSC} as:
\begin{eqnarray}
          \nonumber
  &&     \frac{A_e^{\dag}({\bf r})^{n_r}
               A_e^{\dag}({\bf R})^{n_R}
               A_h^{\dag}({\bf r}_h)^{n_h}
      \tilde{B}_e^{\dag}({\bf R})^k
      \tilde{B}_h^{\dag}({\bf r}_h)^l
             B_e^{\dag}({\bf r})^m |\tilde{0} \rangle}
                {\left(n_r!n_R!n_h!k!l!m! \right)^{1/2}} \\
        \label{basis}
  & & \quad \quad \quad \equiv |n_r n_R n_h ; \widetilde{ k l m} \rangle  \, .
\end{eqnarray}
The tilde sign shows that the transformed
vacuum state $|\tilde{0}\rangle$ and the transformed
operators (\ref{Bog1}) and (\ref{Bog2}) are involved.
In (\ref{basis}) the oscillator quantum number is fixed and equals $k$,
while $M_z= n_r + n_R- n_h -k + l - m$.
The permutational symmetry requires that $n_r-m$ must be even (odd)
for electron singlet $S_e=0$ (triplet $S_e=1$ states).

The new vacuum state $|\tilde{0} \rangle$ is in fact a coherent
$e$--$h$ state (see below). It was shown in Ref.~\onlinecite{SSC}
that it is feasible, though cumbersome,
to calculate the Coulomb matrix elements in
the representation (\ref{basis}).
In this work, we propose a new approach that is based
on the {\em simultaneous\/} diagonalization of the operators
$\hat{k}_{-}$ and
\begin{equation}
        \label{pi+0}
     \hat{\pi}_{+} =
      A_e^{\dag}({\bf r}_1) + A_e^{\dag}({\bf r}_2) - A_h({\bf r}_h) \, .
\end{equation}
Although $\hat{\pi}_{+}$ is not associated with any
exact symmetry, below in Sec.~\ref{sec:unitary} we show
that such an approach reveals new features of the problem
and also leads to great technical simplifications.

\section{Unitary transformation and operator algebra}
                 \label{sec:unitary}

\subsection{Transformation matrix and new coordinates}

The operators  (\ref{Bog1}) and (\ref{Bog2}) have a simple
representation in the new coordinates
$\ro_1 = \sqrt{2} \, {\bf R} - {\bf r}_h$
and
$\ro_2 = \sqrt{2} \, {\bf r}_h - {\bf R}$:
$\tilde{B}_e^{\dag}({\bf R}) =   B_e^{\dag}(\ro_1)$
and
$\tilde{B}_h^{\dag}({\bf r}_h) =  B_h^{\dag}(\ro_2)$.
This transformation can be conveniently expressed in the matrix form:
\begin{equation}
        \label{R-rho}
             \left( \begin{array}{c} \ro_1   \\
                                     \ro_2   \end{array}  \right)
  = \hat{F} \left( \begin{array}{c} {\bf R}  \\
                                    {\bf r}_h  \end{array} \right)
                  \quad , \quad
      \hat{F}= \left( \begin{array}{rr}
            \cosh  \, \Theta & - \sinh  \, \Theta   \\
          - \sinh  \, \Theta &   \cosh  \, \Theta
                       \end{array} \right)  \,
\end{equation}
with $\cosh  \Theta=\sqrt{2}$, $\sinh \Theta=1$.
The matrix $\hat{F}$ is symmetric, $\hat{F}^{\rm T}=\hat{F}$
(T denotes transposition), and unimodular, $|{\rm Det}\hat{F}|=1$,
but non-orthogonal, $\hat{F}^{\rm T} \neq \hat{F}^{-1}$.
The Bose ladder operators are changed under
the Bogoliubov transformations (\ref{Bog1})--(\ref{Bog2})
according to the same representation:
\begin{equation}
        \label{B-Brho}
 \tilde{S} \left( \begin{array}{c}  B^{\dag}_{e}({\bf R}) \\
                   B_{h}({\bf r}_h)  \end{array}  \right) \tilde{S}^{\dag}
 = \hat{F} \left( \begin{array}{c} B^{\dag}_{e}({\bf R}) \\
                                   B_{h}({\bf r}_h)  \end{array} \right)
 =  \left( \begin{array}{c} B^{\dag}_{e}(\ro_1) \\
                                   B_{h}(\ro_2)  \end{array} \right) \, .
\end{equation}
In (\ref{BogS}) we consider real transformation parameters $\Theta$;
generally, $\Theta$ can be complex which corresponds
to the $SU(1,1)$ symmetry. \cite{Klauder}

The Coulomb interparticle interactions (\ref{H_int})
$H_{\rm int}=H_{\rm ee}+H_{\rm eh}$
in the coordinates \{${\bf r},\ro_1,\ro_2$\} take the form
\begin{eqnarray}
        \label{Ham_rho}
H_{\rm ee} &=& \frac{e^2}{\sqrt{2} \epsilon r} \, , \\
        \label{Ham_eh}
H_{\rm eh} &=& - \frac{\sqrt{2}e^2}{\epsilon|\ro_2 - {\bf r}|}
            - \frac{\sqrt{2}e^2}{\epsilon|\ro_2 + {\bf r}|} \, .
\end{eqnarray}
The important result is that $H_{\rm int}$ {\em does not depend\/} on $\ro_1$.
Later on we will see (Sec.~\ref{subs:tr}) that the new coordinates
can be associated with {\em new effective particles\/} in $B$ ---
two electrons with the coordinates ${\bf r}$ and $\ro_1$
and one hole with the coordinate $\ro_2$.

\subsection{Coherent states and Hamiltonian transformation}
                 \label{sec:coherent}

Note that (\ref{basis}) has a mixed form:
the inter-LL operators are expressed in the variables
\{${\bf r},{\bf R},{\bf r}_h$\}, while the intra-LL operators ---
in the variables \{${\bf r},\ro_1,\ro_2$\}.
From (\ref{Ham_rho}) and (\ref{Ham_eh})
it is clear, however, that it is desirable
to work in the coordinates \{${\bf r},\ro_1,\ro_2$\}.
As a first step, let us establish the coordinate representation
of the transformed vacuum $|\tilde{0}\rangle$.
Disentangling the operators \cite{Ezawa,Kirzhnits} in the exponent of
$\tilde{S}$,
one obtains the normal-ordered form
\begin{eqnarray}
  \tilde{S} & = & \exp \left( \tanh \Theta \, B^{\dag}_{h} B^{\dag}_{e}
               \right)     \nonumber \\
    \nonumber
   &\times& \exp \left( -\ln(\cosh  \Theta )
          [B^{\dag}_{e} B_{e} + B^{\dag}_{h} B_{h} +1 ] \right) \\
   &\times& \exp\left( - \tanh \Theta \, B_eB_h \right) \, .
        \label{Sexp}
\end{eqnarray}
Therefore,
\begin{equation}
         \label{vacuum}
 |\tilde{0} \rangle = \tilde{S} |0\rangle = \frac{1}{\cosh  \Theta }
  \exp\left[ \tanh \Theta \,
  B^{\dag}_{h}({\bf r}_h) B^{\dag}_{e}({\bf R}) \right] |0\rangle \, .
\end{equation}
The state (\ref{vacuum}) is a {\em coherent\/} $e$--$h$ state
in the sense that the anomalous two-particle \cite{oneparticle}
expectation value exists,
$\langle \tilde{0}| B^{\dag}_{h}({\bf r}_h)B^{\dag}_{e}({\bf R})
|\tilde{0}\rangle = uv \neq 0$.
In the terminology of quantum optics, \cite{Klauder}
it is a {\em two-mode squeezed state}.
In the present situation of particles in a magnetic field
the squeezing has a {\em direct geometrical meaning}.
In order to see this, let us  obtain a representation of the new vacuum
in the coordinates \{${\bf r},{\bf R},{\bf r}_h$\}.
Using $\tanh \Theta=1/\sqrt{2}$, $\cosh \Theta=\sqrt{2}$, we have
\begin{eqnarray}
         \nonumber
  && \langle {\bf r} {\bf R} {\bf r}_h | \tilde{0} \rangle =
    \frac{1}{\sqrt{2}\,(2\pi l_B^2)^{3/2}} \\
   \label{vac-r}
   &&  \times \exp \left( -\frac{{\bf r}^2 + {\bf R}^2 + {\bf r}_h^2 -
      \sqrt{2}Z^{\ast} z_h }{4 l_B^2} \right) \, .
\end{eqnarray}
Comparing (\ref{vac-r}) with (\ref{zero_LL})
and using $\sqrt{2}Z^{\ast} = z_1^{\ast} + z_2^{\ast}$,
we first note that the new vacuum state $| \tilde{0} \rangle$
contains contributions of an infinite number of $e$-
and $h$- states in the {\em zero\/} LL\@. In fact it is a coherent state
of the hole and the {\em center-of-charge\/} of two electrons, \cite{ee}
and there are correlations in their positions:
$\langle \tilde{0}| {\bf R} \cdot {\bf r}_h | \tilde{0} \rangle =
2\sqrt{2} \, l_B^2 \neq 0$.  It turns out that the probability
distribution function can be presented in the following factored form
\begin{eqnarray}
     \label{vac-prob}
  |\langle {\bf r} {\bf R} {\bf r}_h | \tilde{0} \rangle|^2 &=&
    \frac{1}{2\pi l_B^2} \exp \left( -\frac{{\bf r}^2}{2l_B^2} \right)  \\
  	       \nonumber
 & \times & \frac{2+\sqrt{2}}{4 \pi l_B^2}
        \exp \left[ - \frac{2+\sqrt{2}}{8 l_B^2}
          \left(  {\bf R}-{\bf r}_h \right)^2  \right] \\
         \nonumber
 & \times &  \frac{2-\sqrt{2}}{4 \pi l_B^2}
        \exp \left[ - \frac{2-\sqrt{2}}{8 l_B^2}
          \left( {\bf R}+{\bf r}_h \right)^2  \right] \, .
\end{eqnarray}
This shows that the distribution for the relative coordinate
${\bf R}-{\bf r}_h$ is squeezed {\em at the expense\/} of that for
the coordinate ${\bf R}+{\bf r}_h$, and the variances are
\begin{eqnarray}
	\label{var-}
 \langle \tilde{0}| ({\bf R} - {\bf r}_h)^2 | \tilde{0} \rangle  &=&
   4(2-\sqrt{2}) \, l_B^2 \simeq 2.3 \,  l_B^2        \,  ,       \\
	\label{var+}
  \langle \tilde{0}| ({\bf R} + {\bf r}_h)^2 | \tilde{0} \rangle  &=&
             4(2+\sqrt{2}) \, l_B^2 \simeq 13.7 \, l_B^2  \, .
\end{eqnarray}

  Note now that the representation of $| \tilde{0} \rangle$ in
the new coordinates has a qualitatively different form
\begin{eqnarray}
         \nonumber
  && \langle {\bf r}\ro_1\ro_2| \tilde{0} \rangle =
    \frac{1}{\sqrt{2}\,(2\pi l_B^2)^{3/2}} \\
   \label{vac-ro}
   &&  \times \exp \left( -\frac{{\bf r}^2+\ro_1^2+\ro_2^2+
      \sqrt{2} {\cal Z}_1 {\cal Z}_2^{\ast} }{4 l_B^2} \right) \, ,
\end{eqnarray}
where ${\cal Z}_j= \rho_{jx} +i\rho_{jy}$, $j=1,2$.
It can be seen from (\ref{vac-ro}) that $|\tilde{0} \rangle$
is a coherent state that contains contributions from infinitely many
$e$- and $h$- {\em higher} LL's of the {\em new effective particles}.
This corresponds in fact to an additional unitary transformation
involving the {\em inter\/}-LL ladder operators:
\begin{equation}
   \label{vac-ro2}
 |\tilde{0} \rangle =  \frac{1}{\cosh  \Theta }
  \exp\left[ - \tanh \Theta \,A^{\dag}_{h}(\ro_2) A^{\dag}_{e}(\ro_1)
       \right] |\bar{0} \rangle =  \bar{S}^{\dag} |\bar{0} \rangle \, ,
\end{equation}
where
\begin{eqnarray}
        \label{BogSbar}
   \bar{S} & = & \exp ( \Theta \bar{\cal L} )  \, ,  \\
        \label{genLbar}
  \bar{\cal L}  & = & A^{\dag}_e(\ro_1) A^{\dag}_h(\ro_2)
                        - A_{h}(\ro_2) A_{e}(\ro_1)  \, .
\end{eqnarray}
The new state introduced in (\ref{vac-ro2}),  $| \bar{0} \rangle$,
corresponds to the simultaneous diagonalization
of $\hat{k}_{-}$ and $\hat{\pi}_{+}$,
\begin{equation}
          \label{bar0}
     | 0 \rangle   \stackrel{\hat{k}_{-},\hat{\pi}_{+}}{\longrightarrow}
            |\bar{0} \rangle = \bar{S} \tilde{S} |0 \rangle =
            \bar{S} |\tilde{0} \rangle \, .
\end{equation}
The coordinate representation
\begin{equation}
   \label{vac-ro3}
   \langle {\bf r}\ro_1\ro_2| \bar{0} \rangle =
     \frac{1}{(2\pi l_B^2)^{3/2}}
   \exp \left( -\frac{{\bf r}^2+\ro_1^2+\ro_2^2}{4 l_B^2} \right)
\end{equation}
shows that $| \bar{0} \rangle$ is a {\em true vacuum\/} for both the
intra-LL $B^{\dag}_{e}(\ro_1)$, $B^{\dag}_{h}(\ro_2)$
and inter-LL  $A^{\dag}_{h}(\ro_2)$, $A^{\dag}_{e}(\ro_1)$
operators. The latter transform according to the
representation (\ref{B-Brho}):
\begin{equation}
        \label{A-Arho}
 \bar{S} \left( \begin{array}{c}  A^{\dag}_{e}({\bf R}) \\
                         A_{h}({\bf r}_h)  \end{array}  \right) \bar{S}^{\dag}
 = \hat{F} \left( \begin{array}{c} A^{\dag}_{e}({\bf R}) \\
                                   A_{h}({\bf r}_h)  \end{array} \right)
 =  \left( \begin{array}{c} A^{\dag}_{e}(\ro_1) \\
                                   A_{h}(\ro_2)  \end{array} \right) \, .
\end{equation}
This allows us to perform the desirable complete transformation
$\{ {\bf r},{\bf R},{\bf r}_h \} \rightarrow \{ {\bf r},\ro_1,\ro_2 \}$.
Indeed, using the commutativity $[\bar{S}, A_e^{\dag}({\bf r})]=0$,
the transformation
\begin{eqnarray}
   \nonumber
                A_e^{\dag}({\bf R})^{n_R}
                A_h^{\dag}({\bf r}_h)^{n_h} \bar{S}^{\dag}
 &=& \bar{S}^{\dag} \left( \bar{S}
                       A_e^{\dag}({\bf R})^{n_R}
                       A_h^{\dag}({\bf r}_h)^{n_h} \bar{S}^{\dag} \right)  \\
         \label{Stran}
 &=& \bar{S}^{\dag}
             A_e^{\dag}(\ro_1)^{n_R}
             A_h^{\dag}(\ro_2)^{n_h} \, ,
\end{eqnarray}
and Eqs.~(\ref{basis}) and (\ref{bar0}), we have
\begin{eqnarray}
   \nonumber
 & & | n_{r} n_{R} n_{h}; \widetilde{ k l m} \rangle  = \\
   \nonumber
 & & \frac{ \bar{S}^{\dag}
                   A_e^{\dag}({\bf r})^{n_r}
                   A_e^{\dag}(\ro_1)^{n_R} A_h^{\dag}(\ro_2)^{n_h}
                   B_e^{\dag}(\ro_1)^{k}
                   B_h^{\dag}(\ro_2)^{l}
                   B_e^{\dag}({\bf r})^{m}
   | \bar{0} \rangle}{(n_{r}!n_{R}!n_{h}!k!l!m!)^{1/2}} \\
         \label{bas_new}
 & & \equiv  \bar{S}^{\dag} | \overline{ n_{r} n_{R} n_{h}; k l m} \rangle \, .
\end{eqnarray}
The overline shows that a state is generated in the usual way by the
intra- and inter-LL ladder operators acting on the true vacuum
$|\bar{0}\rangle$ --- all in the representation of
the coordinates $\{{\bf r},\ro_1,\ro_2\}$.

The Hamiltonian $H$ is block-diagonal in the quantum numbers
$k,M_z$ (and $S_e$, $S_h$). Moreover, due to the Landau degeneracy in $k$,
it is sufficient to consider only the states with $k=0$.
This effectively removes one degree of freedom. The constraint
following from conservation of $M_z$ removes another degree of freedom.
This corresponds
to a possible \cite{Simon,Hirsch} partial separation of the CM
motion from internal degrees of freedom for a charged $e$--$h$
system in a magnetic field.

From now on we will consider the $k=0$ states only,
designating such states in (\ref{bas_new}) as
$| \overline{ n_{r} n_{R} n_{h}; l m} \rangle$.
For the Hamiltonian we arrive therefore at the unitary
transformation
\begin{eqnarray}
       \nonumber
 &&\langle \widetilde{ m_2 l_2 }; n_{h2} n_{R2} n_{r2} | H
  | n_{r1} n_{R1} n_{h1}; \widetilde{ l_1 m_1 } \rangle      \\
  \label{me1}
  & &    = \langle \overline{ m_2 l_2 ; n_{h2} n_{R2} n_{r2} } |
               \bar{S} H \bar{S}^{\dag}
  | \overline{n_{r1} n_{R1} n_{h1}; l_1 m_1 } \rangle  \, .
\end{eqnarray}

\subsection{Transformed Hamiltonian}
                \label{subs:tr}

We should now work out how the total Hamiltonian $H=H_0+H_{\rm int}$
changes under the  transformation $\bar{S} H \bar{S}^{\dag}$.
Note first that the free Hamiltonians transform as
$\bar{S}  H_{0e}({\bf r})   \bar{S}^{\dag} = H_{0e}({\bf r})$,
$\bar{S}  H_{0e}({\bf R})   \bar{S}^{\dag} = H_{0e}(\ro_1)$,
and
$\bar{S}  H_{0h}({\bf r}_h) \bar{S}^{\dag} = H_{0h}(\ro_2)$;
this is evident from (\ref{A-Arho}).
Therefore, the transformed free Hamiltonian
$\bar{S} H_0 \bar{S}^{\dag}$ is diagonal in the coordinates
$\{{\bf r},\ro_1,\ro_2 \}$ and describes the aforementioned
new effective particles --- free $e$ and $h$ in a magnetic field.
The peculiarity of the situation is that the whole interaction
Hamiltonian --- {\em before transformation\/}  (\ref{me1}) ---
does not depend on $\ro_1$ at all.

The Hamiltonian of the $e$--$e$ interactions
$H_{\rm ee}=U_{\rm ee}({\sqrt{2}|{\bf r}|})$
can be handled in a straightforward way:
it does not depend on $\ro_1$, $\ro_2$ and, therefore, is invariant
\begin{equation}
     \label{SHee}
    \bar{S} H_{\rm ee} \bar{S}^{\dag} = H_{\rm ee} \, .
\end{equation}
Thus, the matrix elements of the $e$--$e$ interaction
are easily obtained from (\ref{me1}) and (\ref{SHee}):
\begin{eqnarray}
        \label{mat_ee}
 & & \langle \widetilde{ m_2 l_2 }; n_{h2} n_{R2} n_{r2} | H_{\rm ee}
  | n_{r1} n_{R1} n_{h1}; \widetilde{ l_1 m_1 } \rangle \\
     \nonumber
  & & \quad =  \langle \overline{ m_2 l_2 ; n_{h2} n_{R2} n_{r2}} | H_{\rm ee}
    | \overline{ n_{r1} n_{R1} n_{h1}; l_1 m_1 } \rangle   \\
     \nonumber
  & & \quad =  \delta_{n_{R1},n_{R2}}  \delta_{n_{h1},n_{h2}} \delta_{l_1,l_2}
 \delta_{n_{r1}-m_1,n_{r2}-m_2} F_{n_{r1} m_1}^{n_{r2} m_2}  \, ,
\end{eqnarray}
with $F_{n_1 m_1}^{n'_1 m'_1}$ defined as
\begin{eqnarray}
        \label{Uee}
   \nonumber
   \int\! d^2 r \,
    \phi_{n'_1 m'_1}^{(e)*}\left( {\bf r} \right)
         U_{\rm ee}({\sqrt{2}|{\bf r}|})
    \phi_{n_1 m_1}^{(e)}\left( {\bf r} \right)  \\
   = \delta_{n_1-m_1,n'_1-m'_1} F_{n_1 m_1}^{n'_1 m'_1} \, .
\end{eqnarray}
Note the length scale change in $U_{\rm ee}$.
For the Coulomb interaction $U_{\rm ee}(|{\bf r}|)=e^2/\epsilon|{\bf r}|$,
Eq.~(\ref{Uee}) reduces to the matrix elements
$V_{n_1 m_1}^{n'_1 m'_1}$ describing the interaction of the electron
with a fixed negative charge $-e$:
$F_{n_1 m_1}^{n'_1 m'_1}=V_{n_1 m_1}^{n'_1 m'_1}/\sqrt{2}$.
The explicit form of the matrix elements in lowest LL's
can be found elsewhere. \cite{SSC,AHM_86,L&L80}

The Hamiltonian $H_{\rm eh}$ depends on $\ro_2$ and is
therefore affected by the transformation $\bar{S} H_{\rm eh} \bar{S}^{\dag}$.
The generator of the Bogoliubov transformations
$\bar{\cal L}$ (\ref{genLbar})
and $H_{\rm eh}$ do not form a closed algebra of a finite order.
It thus appears that it is not possible to establish the form of
$\bar{S} H_{\rm eh} \bar{S}^{\dag} $ in the general case. \cite{Kirzhnits}
It is possible, however, to determine the form of the matrix elements
of $\bar{S} H_{\rm eh} \bar{S}^{\dag} $ in (\ref{me1}).
Because of the permutational symmetry, the two terms in $H_{\rm eh}$
(\ref{Ham_eh}) give equal contributions; it will be sufficient
to consider the term
$U_{\rm eh}(\ro_2-{\bf r}) = - e^2/\epsilon|\ro_2-{\bf r}|$ only.
In order to illustrate the approach, we consider here the states
in zero LL's
$| \overline{000; l m} \rangle  \equiv |\overline{l m } \rangle$.
Disentangling the operators in $\bar{S}$ analogously to (\ref{Sexp}),
we have
\begin{eqnarray}
    \label{eh00}
& & \quad \langle \overline{m_2 l_2}| \bar{S} U_{\rm eh} \bar{S}^{\dag}
        |\overline{l_1 m_1} \rangle
  \equiv \bar{U}_{0m_1\,0l_1}^{0m_2\,0l_2} \\
       \nonumber
 & & = \case{1}{2} \langle \overline{m_2 l_2}|
  e^{-\case{1}{\sqrt{2}}A_e(\ro_1)A_h(\ro_2)}
           U_{\rm eh}
  e^{-\case{1}{\sqrt{2}}A_h^{\dag}(\ro_2)A_e^{\dag}(\ro_1)}
  |\overline{l_1 m_1} \rangle  \, .
\end{eqnarray}
Expanding the exponents and exploiting the fact that
$U_{\rm eh}(\ro_2-{\bf r})$ does not depend on $\ro_1$, we obtain a series
\begin{equation}
        \label{ser1} 
 \bar{U}_{0m_1\,0l_1}^{0m_2\,0l_2} =
      \case{1}{2} \sum_{p=0}^{\infty}
        \left( \case{1}{2} \right)^p U_{0m_1\,pl_1}^{0m_2\,pl_2} \, .
\end{equation}
The matrix elements of the $e$--$h$ interaction
in different LL's are defined on the wave functions (\ref{WF})
in the usual way as
\begin{eqnarray}
   \nonumber
 & & \int\! d^2 r_1 \int\! d^2 r_2 \,
     \phi_{n'_1 m'_1}^{(e)*}\left( {\bf r}_1 \right)
     \phi_{n'_2 m'_2}^{(h)*}\left({\bf r}_2 \right) \\
   \nonumber
 & &   \times U_{\rm eh}({|{\bf r}_1 - {\bf r}_2|})
    \phi_{n_2 m_2}^{(h)}\left( {\bf r}_2 \right)
    \phi_{n_1 m_1}^{(e)}\left( {\bf r}_1 \right)  \\
        \label{Ueh}
 & & = \delta_{n_1-m_1-n_2+m_2,n'_1-m'_1-n'_2+m'_2}
          U_{n_1m_1\,n_2m_2}^{n'_1m'_1\,n'_2m'_2} \, .
\end{eqnarray}
Note that (\ref{ser1}) includes contributions of the
{\em infinitely many\/} LL's.
The method of performing infinite summations in (\ref{ser1})
is presented in Appendix~\ref{Ap_B}.
For the matrix elements of $H_{\rm eh}$ in zero LL's we
finally have
\begin{eqnarray}
      \label{new-repr}
 & & \langle \overline{m_2 \, l_2}| \bar{S} H_{\rm eh} \bar{S}^{\dag}
                     | \overline{l_1 \, m_1}\rangle   \\
      \nonumber
 & & = \delta_{l_1-m_1,l_2-m_2}\,
       2\sqrt{2}\, \bar{U}_{0m_1\,0l_1}^{0m_2\,0l_2} \, .
\end{eqnarray}
The analytical form of $\bar{U}_{0m_1\,0l_1}^{0m_2\,0l_2}$
is given by Eqs.~(\ref{barU0}) and  (\ref{barU1}).
For the matrix elements of the Hamiltonian $H_{\rm eh}$ involving
higher LL's $n_R$, $n_h$ in (\ref{me1}),
we obtain infinite summations similar to (\ref{ser1}).
These can be performed by the method, which is described
in Appendix~\ref{Ap_B}.

  The mathematical tools developed in this Section allow us to reduce the
three-particle Schr\"odinger equation in $B$ to the secular
equation following from the expansion in the basis (\ref{bas_new}).
Because of $M_z$ and $k$ conservation, we have four
(instead of six) independent orbital quantum numbers.
An important property of this basis is that any
truncation of it does not break the translational invariance,
since the exact quantum number $k$ has been fixed.
    The transformation (\ref{bar0}),  (\ref{me1}),
together with the properties of the transformed Hamiltonian
$\bar{S} H \bar{S}^{\dag}$ established above, are the main formal
results of this paper.

\section{$X^-$ resonances in higher Landau levels}
                 \label{sec:res}
Now we apply the developed formalism for a description
of the $2e$--$h$ states in high magnetic fields
\begin{equation}
      \label{highB}
 \hbar\omega_{\rm ce} \, , \,  \hbar\omega_{\rm ch} \gg
E_0 = \sqrt{\frac{\pi}{2}} \, \frac{e^2}{\epsilon l_B} \, ,
\end{equation}
when LL's remain well-defined; $E_0$ is the characteristic energy
of the Coulomb interactions.
Neglecting mixing between LL's (the high field limit),
the three-particle $2e$--$h$ states can be labeled
by a pair of quantum numbers $(n_en_h)$, describing
the electron LL number $n_e=n_r+n_R$
[see (\ref{me1})] and the hole LL number $n_h$. \cite{SSC}
For the states in, e.g., first electron and zero hole LL's
$(n_en_h)=(10)$,
the basis (\ref{bas_new}) includes the states with
$n_r=1$, $n_R=n_h=0$ and $n_R=1$,
$n_r=n_h=0$ with $l-m+1=M_z$ and $k=0$ fixed.
Even in a given LL the basis is infinite.
Here we present the results of numerical few-particle
calculations of the $2e$--$h$
eigenspectra in several of the lowest LL's $(n_en_h)$ obtained by
diagonalization of finite matrices of order $2-5\times10^2$.
Such finite-size calculations
provide very high accuracy for bound states (discrete spectra)
and are also capable of reproducing
in some detail the structure of the three-particle continuum in high fields.
This is because in our approach
(i) off-diagonal Coulomb matrix elements fall-off exponentially \cite{SSC}
and (ii) the three-particle configurational space in the high field
limit has a dimension of {\em one}:
two exact ($M_z$, $k$) and three approximate ($n_e=n_r+n_R$, $n_h$)
quantum numbers have been fixed.
\begin{figure}[!b]
\epsfxsize=2.8in
\begin{center}
\epsffile{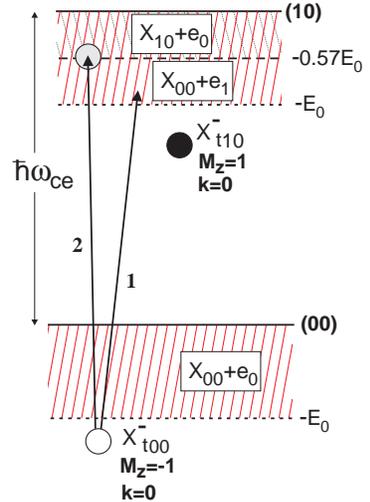}
\end{center}
\caption{Schematic drawing of the $2e$--$h$ electron triplet
$S_e=1$ eigenstates in zero $(n_en_h)$=(00)
and first electron $(n_en_h)$=(10) LL's
and the allowed internal transitions from the
triplet ground state $X^{-}_{t00}$.
The shaded dot indicates the presence within a continuum
of a quasi-bound $X^-$ state (see the text).
}                \label{fig1}
\end{figure}
Schematically, the spectra of the triplet $2e$--$h$ eigenstates
in two lowest LL's are shown in Fig.~\ref{fig1}.
The hatched areas correspond in Fig.~\ref{fig1} to
the three-particle continuum.
It is formed by the states of the neutral magnetoexciton (MX),
which has {\em bound\/} internal $e$--$h$ motion and
{\em extended\/} CM motion, \cite{L&L80}
and an electron in a scattering state; the latter on average is
at infinity from the MX\@.
For the $(n_en_h)=(10)$ LL's, there are two different overlapping MX bands.
One corresponds to the $X_{00}$ MX ($e$ and $h$ in their
zero LL's) plus a second electron in a scattering state in the first LL\@.
The second, narrow continuum corresponds to the $X_{10}$ MX
($e$ the first and $h$ in the zero LL).
The lower continuum edge lies at the $X_{10}$ ground state
energy $E_{10}=-0.574 E_0$, which, for the isolated $X_{10}$,
is achieved at a {\em finite\/} CM momentum
${\bf K}_0 \simeq 1.19 \hbar l_B^{-1}$.
Importantly, this produces in 2D a van Hove singularity
in the $X_{10}$ density of states
$g_{\rm sing}(E) \simeq M_{01}^{1/2} K_0/ \pi \hbar \sqrt{E-E_{10}}$,
where $M_{10} \simeq 3.62 \hbar^2/ E_0 l_B^2 $ is the effective \cite{L&L80}
$X_{10}$ mass.

\begin{figure}[!b]
\epsfxsize=3.4in
\epsffile{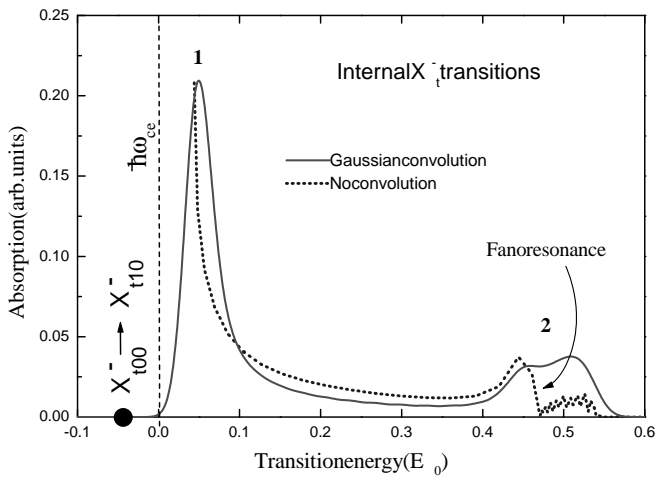}
\vspace*{3ex}
\caption{Dotted line:
spectra of the transitions from the triplet $X^-_{t00}$ ground state
to the continuum in the first electron LL.
Solid line: the spectra convoluted with a Gaussian of the $0.015E_0$ width.
Labeling of peaks corresponds to Fig.~1.
The filled dot shows the position of the forbidden
bound-to-bound $X_{t00}^{-}\rightarrow X_{t10}^{-}$ transition.
}                \label{fig2}
\end{figure}
The bound $X^-$ states form discrete spectra and are
characterized by {\em bound\/} internal motions
of all particles. Such states lie outside the continuum.
In the zero LL's, there is only one bound state --- the triplet
$X_{t00}$ with a small binding energy $0.044E_0$
(counted from the lower continuum edge). \cite{X-th,Dz&S_PRL}
In a 2D system in the high-$B$ limit, there are no bound
singlet $X^-$ states in the zero LL's. \cite{X-th}
In the next electron LL $(n_en_h)=(10)$, there is also only
one bound state, which is the triplet $X^-_{t10}$
with a larger binding energy $0.086E_0$.\cite{SSC,Dz&S_PRL}
We do not discuss here the discrete excited
three-particle states \cite{SSC} that lie above LL's.

In addition, quantum mechanical resonances --- quasi-bound three particle
states --- can exist in the continuum.
Such a possibility appears plausible for charged 2D MX's because of
the van Hove singularities in the density of states of neutral MX's.
Quasi-bound states, because of long-range oscillating tails,
do not have normalizable wave functions. \cite{Baz'}
Nevertheless, they have large probabilities of finding all three particles
together in real space.
In optical transitions quasi-bound states
may produce Fano resonances \cite{Fano} ---
spectra with highly asymmetric line shapes that are
determined by the coupling between a quasi-bound state and a continuum.

We have found spectra of this sort in internal transitions from the 2D
triplet $X^-_{t00}$ ground state to the next electron LL (Fig.~\ref{fig1}).
Such $X^-$ internal transitions are strong and gain strength with $B$.
The transitions must simultaneously satisfy \cite{Dz&S_PRL}
the two exact selection rules: $\Delta k =0$ and $\Delta M_z = \pm 1$.
Because of the selection rules,
the bound-to-bound $X^-_{t00} \rightarrow X^-_{t10}$
transition turns out to be strictly
prohibited in a translationally invariant system. \cite{Dz&S_PRL}
The allowed transitions are therefore photoionizing transitions
to the {\em continuum}. They have intrinsic linewidths $\sim 0.2 E_0$
with a sharp onset at the threshold energy that equals
$\hbar\omega_{\rm ce}$ plus the $X^-_{t00}$ binding energy
(transition~1 in Figs.~1,\,2).
There is also a prominent feature at an energy
about $\hbar\omega_{\rm ce}+0.5E_0$ --- the second peak
denoted as transition~2 in Figs.~1,\,2.
The predicted \cite{Dz&S_PRL} double-peak structure
in the singlet and triplet $X^-$ internal photoionizing
transitions has been observed \cite{X-int} in
quantum wells in magnetic fields. The positions of the peaks are
in good quantitative agreement with calculations performed
for realistic parameters of quasi-2D quantum wells
at finite fields, as has been described in detail
elsewhere. \cite{X-int,Dz&S_PRL}
The existence of the second peak has been previously
associated \cite{Dz&S_PRL} with the high density of
final states at the lower edge of the $X_{10}+e_0$ continuum.

Our present high-accuracy calculations have revealed the fine structure
of the second peak. When the spectra are convoluted with the Gaussian
of the $0.015E_0$ width, which simulates a relatively large inhomogeneous
broadening, the second peak has a ``camel-back'' shape.
However, when no artificial broadening is performed, a shape typical
\cite{Fano} for the Fano {\em antiresonance\/} is clearly present
in the spectra (Fig.~\ref{fig2}). This is an evidence that
quasi-bound charged MX's $X^-$ exist within the
{\em three-particle\/} continuum.
Note that such states are absent in the two-particle spectra
of the strictly 2D neutral MX's that have essentially
bound relative $e$--$h$ motion; \cite{L&L80}
such states exist for bulk \cite{Glutsch} 3D and
confined \cite{1D} quasi-1D neutral MX's.
We considered here the spectra of internal $X^-$ transitions.
Resonances that are optically-active in photoluminescence \cite{X-exp}
are also expected to exist in the spectra.
Experimental search for such resonances require high quality samples
with small inhomogeneous broadening.

\section{Conclusions}
                 \label{sec:concl}

We have developed a novel expansion in LL's that is compatible
with both rotations about the ${\bf B}$ axis and magnetic translations.
The operator approach allows one to partially separate
the center-of-mass from internal degrees of freedom for
charged $e$--$h$ systems in magnetic fields.
The proposed unitary transformation of the Hamiltonian may be
useful in various solid state and atomic physics problems
dealing with systems of charged particles in magnetic fields.
Here we have considered the 2D systems; however, the developed approach
can also be applied to 3D systems \cite{Simon,Hirsch} for
the separation of the coordinates in the plane perpendicular to ${\bf B}$.

We have found evidence that, in addition to discrete bound states,
quasi-bound states (resonances) of charged magnetoexcitons $X^-$
exist in the continuum of higher Landau levels in 2D systems;
this is a qualitatively new feature in the three-particle
spectra in a magnetic field.
Experimentally, such states may be observed as Fano-resonances
in the interband and intraband optical spectra.


The author is grateful to  B.D.\ McCombe for useful discussions.
This work was supported in part by COBASE grant
and by Russian Ministry of Science program ``Nanostructures''.

\appendix

\section{Charged \lowercase{$e$}--\lowercase{$h$} systems
        with an arbitrary number of particles}
                     \label{Ap_A}

Charged $e$--$h$ complexes, such as charged multiple-excitons \cite{X-th}
$X_n^-$ and multiply-charged excitons \cite{Yudson} $X^{-k}$, may be
{\em bound and stable\/} in quasi-2D systems.
Let us demonstrate that for a charged system containing an arbitrary
number of $N_e$ electrons and $N_h$ holes (with, e.g., $N_e>N_h$),
a transformation analogous to (\ref{Bog1})--(\ref{S0}) can also be
performed.
Let us first separate the center-of-masses of the $e$- and $h$- subsystems.
This can be done, for example, with the help of the linear orthogonal
Jacobi transformation: For the electron coordinates we have
$\{ {\bf r}_{ei} \} \rightarrow \{ \tilde{ {\bf r} }_{ei} \}$,
where
$\tilde{ {\bf r} }_{ei}= (\sum_{l=1}^i {\bf r}_{el} - i{\bf r}_{ei+1})/
                                         \sqrt{i(i+1)}$,
$i=1, \ldots , N_e-1$ are the internal coordinates
and $\tilde{ {\bf r} }_{eN_e} \equiv {\bf R}_e  =
\sum_{i=1}^{N_e} {\bf r}_{ei}/\sqrt{N_e}$ is the electron CM coordinate.
The analogous transformation is performed for the hole coordinates.
Note that the orthogonality of the transformations ensures that the
${\bf R}_e$ and ${\bf R}_h$ degrees of freedom carry the charges
$\mp e$, respectively. \cite{Hirsch}
We have therefore
\begin{equation}
       \label{k-eh}
 \hat{k}_{-} = \sqrt{\frac{N_e}{N_e-N_h}} B^{\dag}_{e}({\bf R}_e)
            -  \sqrt{\frac{N_h}{N_e-N_h}} B_{h}({\bf R}_h) \,  ,
\end{equation}
where $B^{\dag}_{e}$ and $B_{h}$ are the $e$- and $h$-CM
intra-LL ladder operators.
We can now see that, analogously to (\ref{Bog1}),
the Bogoliubov transformation diagonalizing
$\hat{\bf k}^2$ should involve the intra-LL $e$- and $h$-
center-of-mass operators
$B^{\dag}_{e}({\bf R}_e)$ and $B_{h}({\bf R}_h)$
with $\Theta = \tanh^{-1}(\sqrt{N_h/N_e})$.

\section{Electron systems}
                     \label{Ap_e}

It is interesting to compare the $e$--$h$ systems with systems of
charges of the same sign (e.g., $e_j < 0$ for all particles $j$).
To illustrate this we consider a simplest possible system
of two negative charges $e_1, e_2 <0$  of masses $m_1$ and $m_2$.
The raising operator $\hat{k}_{-}$ has the form
[cf.\ (\ref{k-}) and (\ref{k-eh})]
\begin{equation}
        \label{k-2e}
 \hat{k}_{-} =   \sqrt{\frac{e_1}{e_1+e_2}} \, B^{\dag}_{e}({\bf r}_1)
              +  \sqrt{\frac{e_2}{e_1+e_2}} \, B^{\dag}_{e}({\bf r}_2) \,  .
\end{equation}
This can be considered to be a result of the unitary transformation
\begin{equation}
        \label{uv2e}
  {\cal S} B^{\dag}_{e}({\bf r}_1) {\cal S}^{\dag} =
 u B^{\dag}_{e}({\bf r}_1)  + v B^{\dag}_{e}({\bf r}_2)
\end{equation}
where ${\cal S}= \exp ( \varphi L )$
with the generator
$L= B^{\dag}_{e}({\bf r}_1) B_{e}({\bf r}_2)
        - B^{\dag}_{e}({\bf r}_2) B_{e}({\bf r}_1)$.
The transformation parameters are given by
$u=\cos \varphi = \sqrt{e_1/(e_1+e_2)}$ and
$v=\sin \varphi = \sqrt{e_2/(e_1+e_2)}$.
Note that contrary to the $e$--$h$ systems [see Eq.~(\ref{vacuum})],
the vacuum state does not change under this transformation:
${\cal S} |0\rangle = |0\rangle$.
Another way of looking at this result is to consider (\ref{uv2e})
as a transformation following from the
{\em orthogonal\/} transformation of the coordinates
\begin{equation}
        \label{R-r}
             \left( \begin{array}{c} {\bf R}_1   \\
                                     {\bf R}_2   \end{array}  \right)
  = \hat{G} \left( \begin{array}{c} {\bf r}_1  \\
                                    {\bf r}_2  \end{array} \right)
                  \quad , \quad
      \hat{G}= \left( \begin{array}{rc}  \cos\varphi   &   \sin\varphi   \\
                                        -\sin\varphi   &   \cos\varphi
                       \end{array} \right)  \, .
\end{equation}
The matrix $\hat{G}$ is orthogonal, i.e., satisfies
$\hat{G}^{\rm T} = \hat{G}^{-1}$.
The electron Bose ladder operators are changed according to
[cf.\ with Eqs.~(\ref{R-rho}) and (\ref{B-Brho})]
\begin{equation}
        \label{B-2e}
 {\cal S} \left( \begin{array}{c}  B^{\dag}_{e}({\bf r}_1) \\
                                   B^{\dag}_{e}({\bf r}_2)
                          \end{array}  \right) {\cal S}^{\dag}
 = \hat{G} \left( \begin{array}{c} B^{\dag}_{e}({\bf r}_1) \\
                                   B^{\dag}_{e}({\bf r}_2)  \end{array} \right)
 =  \left( \begin{array}{c} B^{\dag}_{e}({\bf R}_1) \\
                            B^{\dag}_{e}({\bf R}_2)  \end{array} \right) \, .
\end{equation}
For real parameters of transformation $\varphi$ we deal
with O(2) matrices, in general the symmetry group is $SU(2)$.
The coordinate representation of the vacuum state
$\langle {\bf r}_1 \ldots {\bf r}_j| 0 \rangle \sim
\exp (- \sum_j {\bf r}_j^2/4 l_B^2)$
contains a bilinear form in the exponent and is invariant
under orthogonal transformations.

The orthonormal basis of states with $k=0$ and $M_z=-m$
in, e.g., zero LL is
\begin{equation}
        \label{2ebasis}
  | m \rangle = \frac{1}{\sqrt{m!}} \,
    [u B^{\dag}_{e}({\bf r}_1) - v B^{\dag}_{e}({\bf r}_2)]^m
                  |0\rangle             \, .
\end{equation}
For $u\neq v$ these states do not have definite parity under
the permutation ${\bf r}_1 \leftrightarrow {\bf r}_2$.
The Coulomb interaction energies are given by the expectation values
$\langle m | U_{\rm ee}({\bf r}_1 - {\bf r}_2)  |m \rangle$, which
solves the problem in the high-field limit. Note that the form
of the eigenstates (\ref{2ebasis}) {\em does not depend\/} on the form
of the interaction potential $U_{\rm ee}$
(cf.\ with Ref.~\onlinecite{Pasquier}).
Note that if the charge-to-mass ratio is the same for all particles,
$e_j/m_j= {\rm const}$, the states $(\hat{\pi_{+}})^n | m \rangle$
are also exact eigenstates of the Hamiltonian [see Eq.~(\ref{Pplus})]
and correspond to the {\em free\/} CM motion in the $n$-th LL.

\section{Coulomb matrix elements}
            \label{Ap_B}

In order to perform the infinite summations in (\ref{ser1}),
it is convenient to obtain first the presentation
of the matrix elements (\ref{Ueh}) using the Fourier transform:
Expressing the exponent $\exp(i{\bf q} \!\cdot\! {\bf r})$
in terms of the intra- and inter-LL ladder operators,
one obtains \cite{AHM_86}
\begin{eqnarray}
        \label{UehF}
 &&  U_{n_1 m_1 \, p  m_2}^{n'_1m'_1 \,p m'_2} =
  \int \! \frac{d^2q}{(2\pi)^2} \, \tilde{U}_{\rm eh}(q)  \\
  \nonumber
 && \times
   \langle n'_1 | \hat{D}(i\tilde{q}^{\ast}) |n_1 \rangle    \,
   \langle m'_1 | \hat{D}(\tilde{q})         |m_1 \rangle \\
  \nonumber
 & & \times
   \langle m_2 | \hat{D}(-\tilde{q})         |m'_2 \rangle \,
   \langle p |   \hat{D}(-i\tilde{q}^{\ast}) |p \rangle \, .
\end{eqnarray}
Here $\tilde{q}=(q_x+iq_y)l_B/\sqrt{2}$ and
the matrix elements
of the displacement operator \cite{Klauder}
$\hat{D}(\alpha)=\exp\left(\alpha A^{\dag} - \alpha^{\ast}A \right)$
between the oscillator eigenstates have, e.g., for $n\leq n'$,
the form
\begin{equation}
      \label{dnn1}
 \langle n'|\hat{D}(\alpha)|n \rangle =
  \sqrt{\frac{n!}{n'!}} \, \alpha^{n'-n} \, e^{-\frac{|\alpha|^2}{2}}
     L_n^{n'-n}(|\alpha|^2) \, ,
\end{equation}
where $L_n^s(x)$ are generalized Laguerre polynomials; $L_n^0(x)=L_n(x)$.
Using (\ref{dnn1}) and the generating function of the Laguerre
polynomials \cite{GR}
\begin{equation}
     \label{Gen}
  \sum_{n=0}^{\infty}L_n^{s}(x)z^n=
    \frac{\exp\left(\frac{xz}{z-1}\right)}{(1-z)^{s+1}}
       \, , \hspace{3ex} |z|<1 \, ,
\end{equation}
we obtain
\begin{eqnarray}
       \label{sum_k}
 & & \case{1}{2} \sum_{p=0}^{\infty} \left( \case{1}{2} \right)^p
 \langle p | \hat{D}(-i\tilde{q}^{\ast})
             | p \rangle  \\
         \nonumber
 & & =  e^{- \case{x}{2}} \case{1}{2} \sum_{p=0}^{\infty}
                \left( \case{1}{2} \right)^p  L_p(x) =
                 e^{-\case{3x}{2} }  \, ,
\end{eqnarray}
where $x=q^2l_B^2/2$.
For the matrix elements  (\ref{ser1})
\begin{eqnarray}
        \label{barU0}
&& \bar{U}_{0m_1\,0l_1}^{0m_2\,0l_2} \equiv  \delta_{l_1-m_1,l_2-m_2} \\
      \nonumber
&& \times  \bar{U}_{{\rm min}(m_1,m_2),{\rm min}(l_1,l_2)}(|m_1-m_2|)
\end{eqnarray}
we therefore obtain the integral representation
\begin{eqnarray}
        \label{barU}
   & & \bar{U}_{mn}(s) = \left(\frac{m!n!}{(m+s)!(n+s)!}\right)^{1/2} \\
         \nonumber
   & &  \times   \int \! \frac{d^2q}{(2\pi)^2} \,
          \tilde{U}_{\rm eh}(q) e^{-3x} x^{s+1} \, L_m^s(x) L_n^s(x) \, .
\end{eqnarray}
For the Coulomb interactions with the 2D Fourier transform
$\tilde{U}_{\rm eh}(q)= - 2\pi e^2/\epsilon q$,
the integral in (\ref{barU}) can be calculated analytically
using the generating function (\ref{Gen}),
as has been described in detail elsewhere. \cite{AHM_86,dzyubenko92}
The final result is
\begin{eqnarray}
    \nonumber
  & & \bar{U}_{mn}(s)  =
 - \frac{E_0}{\left[m!(m+s)!n!(n+s)!\right]^{1/2} \, 2^{m+n+s}3^{s+1/2}} \\
   \nonumber
   & & \times  \sum_{k=0}^{m} \, \sum_{l=0}^{n} \,
    C_m^k \, C_n^l \,  \left(\case{2}{3}\right)^{k+l} \,
    [2(k+l+s)-1]!!  \\
    \nonumber
   & & \times  [2(m-k) -1]!!  \sum_{p=0}^{n-l} \,
                C_{k}^{p} \, C_{n-l}^{p} \,  (-1)^p\, p! \,  \\
        \label{barU1}
    & & \times  [2(n-l-p)-1]!! \, .
\end{eqnarray}

\vspace*{-2ex}



\begin{references}

\vspace*{-8ex}


\bibitem[*]{ABD}on leave from General Physics Institute, RAS,
                Moscow 117942, Russia

\bibitem{Kohn}W.~Kohn, Phys. Rev. {\bf 123}, 1242 (1961).

\bibitem{Knox} R.~S.~Knox, {\it Theory of Excitons},
               Solid State Physics, supplement 5
               (Academic Press, New York, 1963).

\bibitem{Brown}J.~Zak,
               Phys. Rev. {\bf 134}, A1602 (1964);
               E.~Brown, in
               {\it Solid State Physics}, V. 22,
               Eds.\ E.~Ehrenreich and F.~Seitz
               (Academic Press, New York, 1968).

\bibitem{G&D}L.~P.~Gor'kov and I.~E.~Dzyaloshinskii,
             JETP {\bf 26}, 449 (1968).

\bibitem{Lamb}W.~E.~Lamb, Phys. Rev. {\bf 85}, 259 (1952).

\bibitem{Simon}J.~E.~Avron, I.~W.~Herbst, and B.~Simon,
               Ann. Phys. (N.Y.) {\bf 114}, 431 (1978).

\bibitem{Hirsch}B.~R.~Johnson, J.~O.~Hirschfelder, and K.~H.~Yang,
                Rev. Mod. Phys. {\bf 55}, 109 (1983).

\bibitem{coll}N.~R.~Cooper and D.~B.~Chklovskii,
             Phys. Rev. B {\bf 55}, 2436 (1997);
             E.~I.~Rashba and M.~D.~Sturge,
             Phys. Rev. B {\bf 63}, 045\,305 (2000);
             A.~B.~Dzyubenko,
             Phys. Rev. B {\bf 64},  241101 (R) (2001)
             (cond-mat/0106229).

\bibitem{Girvin}S.~M.~Girvin and T.~Jach,
             Phys. Rev. B {\bf 29}, 5617 (1984).

\bibitem{Ezawa}Z.~F.~Ezawa, {\it Quantum Hall Effects\/}
               (World Scientific, Singapore, 2000).

\bibitem{Pasquier}V.~Pasquier,
             Phys. Lett. B {\bf 490}, 258 (2000).

\bibitem{X-th}J.~J.~Palacios, D.~Yoshioka, and A.~H.~MacDonald,
             Phys. Rev. B {\bf 54}, R2296 (1996);
           A.~W\'ojs,  J.~J.~Quinn, and P.~Hawrylak,
           Phys. Rev. B {\bf 62}, 4630 (2000);
              C.~Riva, F.~M.~Peeters, and K.~Varga,
              Phys. Rev. B {\bf 63}, 115\,302 (2001).

\bibitem{SSC} A.~B.~Dzyubenko, Solid State Commun. {\bf 113}, 683 (2000).

\bibitem{X-exp}See, e.g.,
            K.~Kheng, R.~T.~Cox, Y.~Merle d'Aubigne,
            F.~Bassani, K.~Saminadayar, and S.~Tatarenko,
            Phys. Rev. Lett. {\bf 71}, 1752 (1993);
               A.~J.~Shields, M.~Pepper, M.~Y.~Simmons, and D.~A.~Ritchie
               Phys. Rev. B {\bf 52}, 7841 (1995);
          S.~Glasberg,  G.~Finkelstein, H.~Shtrikman, and I.~Bar-Joseph,
          Phys. Rev. B {\bf 59}, R10\,425 (1999)
               and references therein.

\bibitem{Dz&S_PRL}A.~B.~Dzyubenko and A.~Yu.~Sivachenko,
                  Phys. Rev. Lett. {\bf 84}, 4429 (2000).

\bibitem{Fano}U.~Fano, Phys. Rev. {\bf 124}, 1866 (1961).

\bibitem{AHM_86} A.~H.~MacDonald and D.~S.~Ritchie,
                 Phys. Rev. B {\bf 33}, 8336 (1986).

\bibitem{Kirzhnits}D.~A.~Kirzhnits, {\it Field Theoretical Methods in
             Many-Body Systems\/} (Pergamon Press, Oxford, 1967), p.\ 372.


\bibitem{Klauder}J.~R.~Clauder and B.-S.~Skagerstam,
                {\it Coherent States. Application in Physics
                 and Mathematical Physics\/}
                (World Scientific, Singapore, 1985);
	           W.-M. Zhang, D.~H.~Feng, and R.~Gilmore,
                   Rev. Mod. Phys. {\bf 62}, 867 (1990).


\bibitem{oneparticle}For work on single-particle coherent and
                squeezed states in magnetic fields see
	        I.~A.~Malkin and V.~I.~Man'ko,
                JETP {\bf 28}, 527 (1969);
		A.~Feldman and A.~H.~Kahn,
                Phys. Rev. B {\bf 1}, 4584 (1970);
		E.~I.~Rashba, L.~E.~Zhukov, and A.~L.~Efros,
                Phys. Rev. B {\bf 55}, 5306 (1997);
		M.~Ozana and A.~L.~Shelankov,
		Fiz. Tverd. Tela {\bf 40}, 1405 (1998)
                [Phys. Solid State {\bf 40}, 1276 (1998)]
		and references therein.

\bibitem{ee}For a system of charges of the same sign (e.g., $e_j<0$)
            a diagonalization of the operator $\hat{k}_{-}$
            does not produce a coherent state as a new vacuum state
            (see Appendix~\protect\ref{Ap_e}).

\bibitem{L&L80}I.~V.~Lerner and Yu.~E.~Lozovik,
               JETP {\bf 51}, 588 (1980).

\bibitem{Baz'}A.~I.~Baz', Ya.~B.~Zel'dovich, and A.~M.~Perelomov,
               {\it Scattering, Reactions and Decay in Non-Relativistic
                     Quantum Mechanics}
               (Israel Program for Scientific Translations,
                 Jerusalem, 1969).

\bibitem{X-int}H.~A.~Nickel, G.~S.~Herold,
    T.~Yeo, G.~Kioseoglou, Z.~X.~Zhiang, B.~D.~McCombe, A.~Petrou,
    D.~Broido, and W.~Schaff,
    Phys. Stat. Sol. B {\bf 210}, 341 (1998);
             A.~B.~Dzyubenko, A.~Yu.~Sivachenko,
             H.~A.~Nickel, T.~M.~Yeo, G.~Kioseoglou, B.~D.~McCombe,
             and A.~Petrou,
             Physica E {\bf 6}, 156 (2000).

\bibitem{Glutsch} S.~Glutsch, U.~Siegner, M.-A.~Mycek, and D.~S.~Chemla,
                  Phys. Rev. B {\bf 50}, 17\,009 (1994);
                  U.~Siegner, M.-A.~Mycek, S.~Glutsch, and D.~S.~Chemla,
                  Phys. Rev. B {\bf 51}, 4953 (1995).

\bibitem{1D} M.~Graf, P.~Vogl, and A.~B.~Dzyubenko,
             Phys. Rev. B {\bf 54}, 17\,003 (1996).

\bibitem{Yudson} V.~I.~Yudson,
                 Phys. Rev. Lett. {\bf 77}, 1564 (1996).

\bibitem{GR} I.~S.~Gradshtein, I.~M.~Ryzhik,
             {\it Table of Integrals, Series, and Products\/}
             (Academic, San Diego, 1980).

\bibitem{dzyubenko92}A.~B.~Dzyubenko,
                     Sov. Phys. Solid State {\bf 34}, 1732 (1992).


\end{references}
\end{document}